\documentclass[12pt]{article}
\usepackage{latexsym}
\usepackage{amsfonts,color}
\usepackage{amssymb,amsmath,amsthm}
\usepackage{epsfig}
\usepackage{graphicx}

\title{An integrable hierarchy with a 
perturbed 
H\'{e}non-Heiles system}

\author{
A.N.W.~Hone
\thanks{Institute of Mathematics, Statistics \&
Actuarial Science (IMSAS),
University of Kent, Canterbury CT2 7NF, UK.
E-mail: anwh@kent.ac.uk}, V. Novikov \thanks{IMSAS, Kent, as above.  
E-mail: vn9@kent.ac.uk}   
and C. Verhoeven \thanks{Dienst Theoretische Natuurkunde, Vrije
Universiteit Brussel, Pleinlaan 2, B-1050 Brussels, Belgium.
E-mail: cverhoev@vub.ac.be}
}

\begin{document}

\renewcommand{\theequation}{\arabic{section}.\arabic{equation}}

\newcommand{\beq}{\begin{equation}}

\newcommand{\eeq}{\end{equation}}

\newcommand{\bea}{\begin{eqnarray}}

\newcommand{\eea}{\end{eqnarray}}

\newcommand{\lop}{{\mathcal{L}}} 
\newcommand{\aop}{{\mathcal{A}}} 

\newcommand\F{{\bf F}}

\newcommand\G{{\bf G}}

\newcommand\dd{\mathrm{d}}
\newcommand\LM{{\bf L}}

\newcommand\M{{\bf M}}

\newcommand\N{{\bf N}}
\newcommand\A{{\bf A}}
\newcommand\B{{\bf B}}
\newcommand\C{{\bf C}}
\newcommand\D{{\bf D}}
\newcommand\E{{\bf E}}
\newcommand\J{{\bf J}}
\newcommand\K{{\bf K}}
\newcommand\R{{\bf R}}

\newcommand\la{{\lambda}}

\newcommand\om{{\omega}}

\newcommand\ka{{\kappa}}
\newcommand\al{{\alpha}}

\newcommand\be{{\beta}}

\newcommand\ga{{\gamma}}

\newcommand\de{{\delta}}

\newcommand\ze{{\zeta}}

\newcommand\PS{{\mathcal P}}

\newcommand\larra{{\longrightarrow}}

\newcommand\TLM{{\bf \tilde{L}}}

\newcommand\TN{{\bf \tilde{N}}}

\newcommand\tq{{\tilde{q}}}

\newcommand\tp{{\tilde{p}}}

\newcommand\tz{{Tzitzeica$\,\,$}}
\newcommand\eq{{equation$\,\,$}}

\newcommand\rd{{\mathrm{d}}}

\maketitle

\begin{abstract} 
We consider an integrable scalar partial differential equation 
(PDE) that is second order in time. By rewriting it as a system and 
applying the Wahlquist-Estabrook prolongation algebra method, we obtain 
the zero curvature representation of the equation, which leads to a 
Lax representation in terms of an 
energy-dependent Schr\"{o}dinger spectral problem of the  
type studied  by Antonowicz and Fordy. The solutions of this PDE system, 
and of its associated hierarchy of commuting flows, display weak Painlev\'e 
behaviour, i.e. they have algebraic branching. By considering the 
travelling wave solutions of the next flow in the hierarchy, we find an 
integrable perturbation of the case (ii) H\'{e}non-Heiles system  
which has the weak Painlev\'{e} property. We perform 
separation of variables  for this generalized   
H\'{e}non-Heiles system, and describe the corresponding solutions 
of the PDE.  
\end{abstract}

\section{Introduction}

Recently one of us has been classifying integrable 1+1-dimensional 
scalar partial differential 
equations (PDEs) that are second order in time \cite{mnw}, including 
those of the type 
\beq 
\label{type} 
w_{tt}=w_{nx,t}+F[w,w_t, w_x,w_{xt},\ldots] 
\eeq 
(where $F$ is a nonlinear function of $w$ and its $x$ and $t$ derivatives). 
This large class of scalar PDEs includes the Boussinesq equation  
$$ 
u_{tt}=u_{4x}+ (u^2)_{xx}, 
$$ 
a well known integrable equation in shallow water wave theory.  
In the course of performing the classification, the equation 
\beq 
\label{tteq} 
w_{tt}=w_{3x,t}+8w_xw_{xt}+4w_{xx}w_t-2w_xw_{4x}-4w_{xx}w_{3x} 
-24w_x^2w_{xx}. 
\eeq 
was found. The above equation was obtained  
by means of the perturbative symmetry approach to the classification 
of integrable PDEs, as presented in \cite{miknov}. In the 
latter approach, the existence of infinitely many commuting symmetries 
is taken as the defining property of an integrable PDE, and by 
representing a  PDE such as (\ref{type}) 
symbolically (in Fourier space) one obtains 
an infinite sequence of necessary 
conditions for integrability. The equation (\ref{tteq}) was 
isolated by the requirement that it should satisfy the first few of these 
conditions. 

Obviously, checking a finite number of necessary conditions 
does not prove integrability. However, 
in practice it can be proved {\it a posteriori} 
that only the first few conditions are sufficient for integrability 
within a given class of equations.  
In any case, having obtained a particular 
equation such as 
(\ref{tteq}) in the course of classifying the general class 
of PDEs (\ref{type}), 
its integrability must then be proved constructively, either by demonstrating 
that it is explicitly linearizable, 
or by showing that it can be derived as the compatibility 
condition for a linear system (Lax pair) of the form 
\beq 
\label{laxp} 
\lop\psi =0, \qquad \psi_t=\aop\psi, 
\eeq 
for suitable operators $\lop$, $\aop$ having coefficients dependent  
on $w$ and its derivatives, and on  a spectral parameter $\lambda$. 
Once a Lax pair 
(or, alternatively, a linearization) is known, then the infinite 
hierarchy of symmetries of an integrable PDE can be obtained directly.  
To look for a Lax pair, it is most convenient to try to obtain 
the PDE (\ref{tteq}) in the form  of a zero curvature equation 
 \beq \label{zc}  
\F_t-\G_x+ [\, \F , \G \,]=0, 
\eeq 
this being the compatibility
condition for the matrix linear system 
\beq \label{linz} 
\Psi_x=\F\, \Psi, \qquad  
\Psi_t=\G \, \Psi  
\eeq 
(where the matrices $\F$, $\G$ are functions of $u$ and its derivatives, 
and of a spectral parameter).  

In the next section we apply the 
Wahlquist-Estabrook prolongation algebra method 
to the PDE (\ref{tteq}) in order to obtain a zero 
curvature representation for it. The calculations are much 
facilitated by setting $u=w_x$ and $v=w_t-w_{3x}-3w_x^2$ 
and rewriting the scalar equation (\ref{tteq})  
as a  two-component system (see (\ref{sys}) below).  
We find that the simplest zero 
curvature   
representation for the system (\ref{sys}) 
has $\F ,\G \in \mathrm{sl}(2)$, and this is 
equivalent to a scalar Lax pair (\ref{laxp}) 
with the Lax operator being 
the energy-dependent Schr\"odinger operator 
 \beq \label{lop} 
\lop = \lambda \, \partial_x^2 + \left(\frac{v}{4}+u\lambda - \lambda^2\right), 
\eeq 
which is an example of the general type 
\beq 
\label{eschro} 
\lop =\sum_{j=0}^N \lambda^j(\epsilon_j \, \partial_x^2 +u_j) 
\eeq 
studied by Antonowicz and 
Fordy \cite{af1, af}. Some particular 
examples of energy-dependent Schr\"odinger 
hierarchies were considered earlier by Jaulent and Miodek \cite{jm}  
and 
Martinez Alonso \cite{mal} 
(see also the more 
recent work of Shabat and Martinez Alonso \cite{malshabat}). 
The Lax pair for (\ref{tteq}) allows us to write down its 
hierarchy of higher symmetries. Subsequent sections concern the 
weak Painlev\'e property for the PDE (section 3), and travelling wave 
solutions of members of the hierarchy (section 4). It turns 
out that the stationary flow of the next member of the 
hierarchy is equivalent to an integrable perturbation of the 
case (ii) 
H\'enon-Heiles system; this extends a result of Fordy 
concerning the fifth order KdV equation \cite{fordyhh}. 
We briefly remark on similarity reductions in our conclusions 
(section 5).

\section{Prolongation algebra and integrable hierarchy}  

\setcounter{equation}{0}

A Lax pair for  the scalar PDE (\ref{tteq}) can be found in 
a straighforward way 
by first writing it in the form of a two-component system, that is 
\beq
\label{sys}
\begin{array}{ccl}
u_t & = & u_{xxx}+6uu_x+v_x, \\
v_t & = & 4u_xv+2uv_x,
\end{array}
\eeq
in which form the reduction $v=0$ to the KdV equation
\beq \label{kdv}
u_t  =  u_{xxx}+6uu_x
\eeq
is made manifest.
Starting from the above system, we then proceed to  
apply the prolongation algebra method 
due to Wahlquist
and Estabrook \cite{we} (see also 
\cite{fordy} for an excellent exposition). 

We 
start by assuming that the system (\ref{sys}) has a 
zero curvature representation (\ref{zc}) where the matrix 
$\F$ depends only on $u,v$, so  $\F =\F (u,v)$, and $\G$ depends 
on $v$ and derivatives of $u$ of order at most two, so   
$\G =\G (u,u_x,u_{xx},v)$ (with the dependence on a 
spectral parameter being suppressed). Plugging this ansatz into the 
zero curvature equation and substituting for $u_t$ and $v_t$ 
from (\ref{sys}) yields 
\beq 
\label{zcons} 
\F_u\,(u_{3x}+6uu_x+v_x)+F_v\,(4u_xv+2uv_x) 
-\G_v\,v_x-\sum_{j=0}^2\G_{u_{jx}}\,u_{(j+1)x}+[\F , \G ]=0       
\eeq  
(where subscripts denote partial derivatives). Reading off 
the coefficient of $u_{3x}$ in (\ref{zcons}), we have immediately 
that 
$$ 
\G_{u_{xx}}=\F_u, 
$$ 
so that upon integrating we find 
\beq \label{gexpr} 
\G (u,u_x,u_{xx},v)=\F_u\, u_{xx}+\hat{\G}(u,u_x,v)  
\eeq 
for some matrix function $\hat{\G}$. 
Substituting back for $\G$ from (\ref{gexpr})  
into (\ref{zcons}), we can then compare coefficients of $u_{xx}$, and 
we can continue in this way until we obtain explicit expressions 
for $\F$ and $\G$ as differential polynomials in $u,v$ and their 
derivatives with matrix  coefficients. These constant coefficients must further 
satisfy certain algebraic relations, including commutators, and for 
the Lax pair to exist it is necessary that there should be a non-trivial 
realization of these relations within a suitable Lie algebra. With the 
substitution of (\ref{gexpr}) and subsequent integrations and 
back substitutions into (\ref{zcons}), we find that $\F$ takes 
the form 
$$ 
\F =\A u^2 +\B u +\C v + \D, 
$$ 
where the matrices $\A , \B , \C , \D$ are constants. However, we 
note that from the relations between $\A$ and the other 
coefficients of $\F$ and $\G$, we can consistently set $\A =0$, so 
that $\F$ is linear in $u$ and $v$. Having made this simplification, we 
find that $\G$ has the form 
$$ 
\G = \B u_{xx} + [\D , \B ] u_x +(\B +2u \C )v+(6\B -[\B ,[\B , \D ]]) 
\frac{u^2}{2} +[\D ,[\D , \B ]]u+\E. 
$$ 
The relations that must be satisfied by the coefficients are as follows: 
\beq \label{c1} 
[\B , \C ]=0, \qquad 
2\C +[\C , [\D , \B ]]=0, \qquad 
({\tt ad} \, \B )^3 \, \D =0; 
\eeq 
\beq 
\label{c2} 
[\C , ({\tt ad} \, \B )^2 \, \D ]=0, 
\quad 
[\B , ({\tt ad} \, \D )^2 \, \B ] +\frac{1}{2}[\D , 
6\B - ({\tt ad} \, \B )^2 \, \D ] = 0;  
\eeq 
\beq \label{c3}
2[\D , \C ]+ [\C , ({\tt ad} \, \D )^2 \, \B ] =0, 
\quad 
[\B , \E ]+ ({\tt ad} \, \D )^3 \, \B =0; 
\eeq 
\beq \label{c4} 
[\D , \B ]+ [\C , \E ]=0, \qquad [\D ,\E ]=0. 
\eeq 
 
We now seek to realize the above relations non-trivially 
in a suitable Lie algebra. 
To begin with we suppose that  
\beq \label{cdef} 
\C =\xi \B 
\eeq 
for some scalar $\xi$, so that the first relation (\ref{c1}) 
is satisfied, while the second one implies that  
\beq 
[\K ,\B ]=2\B, 
\label{cartan} 
\eeq 
where we have introduced 
\beq 
\K = [\D , \B]. 
\label{kdef} 
\eeq 
The relation (\ref{cartan}) already suggests that 
$\B$ and $\K$ should be two of the three basis elements of an sl(2) 
subalgebra. 

If we now consider the third relation in (\ref{c1}) 
then we see that 
$$ 
({\tt ad} \, \B )^3  \D =-({\tt ad} \, \B )^2 \K =[\B ,2\B ]=0, 
$$  
using (\ref{cartan}), so this relation is automatically satisfied 
given the assumption (\ref{cdef}). Furthermore, using (\ref{cdef}) 
and (\ref{cartan}) we see that both relations (\ref{c2}) and 
the first relation (\ref{c3}) also hold automatically. Let us 
now consider the first equations in (\ref{c4}), which becomes 
$$ 
[\B ,\xi \E -\D ]=0. 
$$ 
The above obviously holds if we choose to set 
$$
\E= \xi^{-1}\D, 
$$ 
in which case the second relation (\ref{c4}) is trivially true.   
Finally, we have the second relation (\ref{c3}) still to be satified, 
which becomes 
\beq \label{finc} 
-\xi^{-1}\K + [\D ,[\D , \K ]]=0. 
\eeq 
Adding a multiple of $\B$ to $\D$ does not change the 
commutation relation (\ref{kdef}), so we set 
$$ 
\D = -\J -\la \B 
$$ for some constant $\la$, and then we find it is consistent to  
take $\J$ and $\B$ as the Chevalley generators of an sl(2)  
algebra, so that  
$$ 
[\B , \J ]=\K , \qquad 
[\K, \B ]=2\B , \qquad 
[\K , \J ]=-2\J , 
$$ 
and the relation (\ref{finc}) is satisfied 
provided that the constants $\xi$ and $\la $ are related 
according to 
$$ 
\xi =\frac{1}{4\la }.  
$$  

With the above choices we see that the final expressions for $\F$ 
and $\G$ take the form 
$$ 
\F=\left(u-\la +\frac{v}{4\la}\right)\, \B -\J 
$$ 
and 
$$ 
\G = \left( u_{xx}+2u^2+v +2u\la -4\la^2 +\frac{uv}{2\la} 
\right) \B 
+u_x \K -(2u+4\la )\J .      
$$
If we then choose 
the standard matrix representation of sl(2) then $\F$ is just  
$$ 
\F=\left(\begin{array}{cc} 0 & u-\la +\frac{v}{4\la} \\ 
-1 & 0 \end{array} \right), 
$$ 
and then setting $\Psi = (-\psi_x,\psi)^T$ we see 
that the $x$ part of the matrix linear system (\ref{linz}) 
is equivalent to the energy-dependent Schr\"{o}dinger 
equation  
\beq 
\label{schrod} 
\frac{1}{\la}\lop \, \psi\equiv \left(\partial_x^2 + 
\frac{v}{4\la}+u-\la \right)\,\psi =0  
\eeq 
for $\psi$, with the operator $\lop$ as 
in (\ref{lop}). Similarly, the $t$ part 
of (\ref{linz}) implies that the time derivative of $\psi$ is 
given by 
\beq \label{psit} 
\partial_t \,\psi=(2u+4\la )\psi_x -u_x\psi. 
\eeq 
Very recently, we learnt that 
Shabat also derived the system (\ref{sys}) as a reduction 
of the so called universal solitonic hierarchy; see the 
unnumbered equation half way down p. 622 in \cite{shabat}.

The construction of integrable hierarchies associated 
with energy-dependent Schr\"odinger operators of the general 
type (\ref{eschro}) 
was described 
in detail by Antonowicz and Fordy \cite{af0, af1, af}.   We now  
present the explicit form of their construction for the particular 
linear problem (\ref{schrod}), showing how this leads 
to the hierarchy of commuting flows for the system (\ref{sys}), 
that correspond to the symmetries of the scalar PDE (\ref{tteq}). 
Starting from the Schr\"odinger equation (\ref{schrod}), the 
sequence of compatible linear evolution equations for the wave function 
$\psi$  takes the form 
\beq \label{psitn} 
\partial_{t_{2N+1}}\, \psi = \left( 2 P_N \partial_x - P_{N,x} 
\right)\,\psi 
\equiv \aop\,\psi  
\eeq 
for integers $N=0,1,2,\ldots$ that label the flows, 
where $P_N$ are certain polynomials  
in $\la$ whose coefficients are functions of $u$, $v$ and their 
derivatives. For each $N$, the requirement of 
compatibility between the time evolution 
(\ref{psitn}) and the Schr\"odinger equation (\ref{schrod}) 
leads to the 
generalized Lax equation 
\beq 
\label{glax} 
\partial_{t_{2N+1}}\lop =[\aop ,\lop ]+ 4 P_{N,x} \cdot \lop , 
\eeq 
which means that 
the potential of the Schr\"odinger 
operator $\lop$ must evolve according to  
\beq \label{potn} 
\partial_{t_{2N+1}}W=(\partial_x^3+4W\partial_x+2W_x)\, P_N, 
\qquad 
W= \frac{v}{4\la}+u-\la . 
\eeq 

For each $N$, the equation (\ref{potn}) encodes the 
evolution of $u$ and $v$ with respect to the time $t_{2N+1}$, 
but the precise from of each $P_N$ 
must be specified 
in order to obtain the explicit form of these evolution equations. 
In fact a compatible sequence of polynomials $P_N$, of  
degree $N$ in $\la$ for each integer $N\geq 0$,  
can be consistently generated by considering the product 
$$ 
\PS =\psi \psi^\dagger 
$$ 
of two independent solutions of the Schr\"odinger equation 
(\ref{schrod}). It is well known \cite{hone} that $\PS$ satisfies 
\beq 
\label{ep} 
\PS \PS_{xx}-\frac{1}{2}\PS_x^2+2W\PS^2 + 8\la =0,  
\eeq 
which is known as the 
Ermakov-Pinney equation \cite{erm, pin};  
note that we have fixed the Wronskian 
$$\psi_x \psi^\dagger -\psi \psi^\dagger_x=4\sqrt{\la }. $$     
It is also 
useful to observe that the differential consequence of (\ref{ep}) is  
a third order linear equation for $\PS$, namely 
\beq \label{plin} 
(\partial_x^3+4W\partial_x+2W_x) \, \PS =0
\eeq 
We can expand the solution of (\ref{ep}) about $\la=\infty$, so 
that 
\beq 
\label{expan} 
\PS = \sum_{j=0}^\infty f_j\la^{-j}, \qquad f_0=2, 
\eeq 
and,  having fixed the sign of the leading coefficient 
$f_0$, we find that all subsequent coefficients $f_j$ are 
determined uniquely from the Ermakov-Pinney equation by 
a recursion of the form 
$f_j=(\, \mathrm{differential \, polynomial \, in} \, 
u,v,f_k, \,\, k<j\, )=(\, \mathrm{differential \, polynomial \, in} \, 
u,v\, )$.  
Then $\PS$ serves as a generating function for the 
sequence of polynomials $P_N$ in the sense that 
\beq 
\label{polys} 
P_N= (\la^N \PS )_+ = \sum_{j=0}^N  f_j \la^{N-j} 
\eeq  
The first few of the $f_j$  
are $$f_0=2, \qquad  f_1=u , \qquad  f_2=(u_{xx}+3u^2+v)/4.$$

In order to write down the hierarchy of flows (\ref{potn}) 
explicitly in Hamiltonian form, it is helpful 
to observe that from (\ref{plin}) the coefficients of $\PS$ satisfy 
the recursion 
$$ 
4\partial_x f_{j+1}=(\partial_x^3+4u\partial_x+2u_x)\, f_j+ 
\frac{1}{2}\Big( 2v\partial_x+v_x\Big)\, f_{j-1}     
$$ 
for $j=0,1,2,\ldots$. Thus,  
upon expanding out in powers of $\la$, all but the two lowest 
orders in (\ref{potn}) cancel out, leaving the equations 
\beq 
\label{biham} 
\partial_{t_{2N+1}}\left(\begin{array}{c} 
u \\ 
v \end{array} \right) = {\bf B}_0  
\left(\begin{array}{c} 
\delta_u \mathcal{H}_{N+1} \\ 
\delta_v \mathcal{H}_{N+1}
\end{array} \right) = 
{\bf B}_1     
\left(\begin{array}{c} 
\delta_u \mathcal{H}_{N} \\ 
\delta_v \mathcal{H}_{N}
\end{array} \right), 
\eeq 
where 
\beq \label{hops}  
{\bf B}_0 = \left(\begin{array}{cc}
4\partial_x & 0  \\ 
0 & 16v\partial_x +8v_x  
\end{array}\right), \quad  
{\bf B}_1 = \left(\begin{array}{cc}
\partial_x^3+4u\partial_x+2u_x & 4v\partial_x +2v_x \\ 
4v\partial_x +2v_x & 0 
\end{array}\right),  
\eeq 
and 
\beq \label{hamders} 
\delta_u \mathcal{H}_N = 4\delta_v \mathcal{H}_{N+1} = f_N, 
\eeq 
with e.g.  
$$ 
\mathcal{H}_0=2\int u\, dx, \quad 
\mathcal{H}_1=\frac{1}{2}\int (u^2+v)\, dx, \quad 
\mathcal{H}_2=\frac{1}{8}\int (-u_x^2+2u^3+2uv)\, dx. 
$$ 
These conserved quantities are just 
two-field generalizations of the well 
known Hamiltonians for the KdV hierarchy, to which they 
reduce upon setting $v=0$.   
By the results of Antonowicz and Fordy concerning energy-dependent 
Schr\"odinger hierarchies \cite{af0, af1, af}, ${\bf B}_0$ and 
${\bf B}_1$ form a compatible pair of Hamiltonian operators, and 
the existence of the  sequence of functionals 
$\mathcal{H}_N$ satisfying (\ref{hamders}) is also 
guaranteed (see e.g. the Lemma on p.468 of \cite{af}). 

Thus we see that (\ref{biham}) constitutes a bi-Hamiltonian 
formulation of the integrable hierarchy whose first 
non-trivial member is the system (\ref{sys}). 
From (\ref{hamders}) it is clear that 
the differential polynomials $f_j$ are gradients of conserved functionals,  
and the 
system (\ref{sys}) just corresponds to the $N=1$ 
flow with $t_3=t$. 
We shall return to this 
formulation of the flows 
in section 4 when we consider travelling wave solutions.  
   
\section{Weak Painlev\'{e} property} 
 
\setcounter{equation}{0}

The connection between the singularity structure 
of differential equations and their integrability has been 
exploited since at least the time of Kowalevksi \cite{kow}. 
Some time later, 
this connection led 
Ablowitz, Ramani and Segur 
to make their conjecture \cite{ars} that (up to possible changes 
of variables) all ordinary differential equations 
arising as reductions of integrable PDEs should have the 
Painlev\'e property, i.e. that all movable singularities 
of their general solution should be poles. 
While testing for the Painlev\'e property is 
a very useful tool for isolating integrable ODEs and PDEs, 
it has long been known that insisting on only pole 
singularities is too strong a requirement for integrability. Indeed, 
there are many examples of integrable systems which have the weaker 
property that all movable singularities of the general solution have 
only a finite number of branches, and this {\it weak} 
Painlev\'e property was proposed by Ramani et al. \cite{weak} 
as a more general criterion for integrability. 

In finite-dimensional classical Hamiltonian mechanics  
there are many examples of Liouville integrable systems 
with the weak Painlev\'e property (see e.g. \cite{abenda, abf}), 
including the classical problem 
of geodesic motion on ellipsoids that was solved by Jacobi \cite{jacobi}.    
In this section we apply the weak Painlev\'{e} test of \cite{weak} 
to the PDE system (\ref{sys}), before making some general remarks 
concerning weak Painlev\'e expansions  
for integrable hierarchies derived from energy-dependent Schr\"odinger 
operators, some of which were considered in \cite{honeed}. 
The ODE reductions of the system (\ref{sys}) 
that are considered in the next section 
provide further examples of integrable finite-dimensional systems 
with the weak Painlev\'e property.

For the system (\ref{sys}) there are two types of possible 
expansion around movable singularities. The first type of expansion around a 
movable (non-characteristic) 
singular manifold $\phi (x,t)=0$ has 
algebraic branching, and the leading order terms in this 
principal balance have the 
form 
\beq 
\label{prinb} 
u\sim \phi_t/(2\phi_x)+ a\phi^{2/3}, 
\qquad 
v\sim \frac{2a(\phi_x)^2}{9\phi^{4/3}}, 
\eeq 
where 
$a=a(x,t)$ 
is an arbitrary function.  
To find the resonances, where new arbitrary functions appear 
in the expansion, we substitute 
$$ 
u\sim \phi_t/(2\phi_x)+ a \phi^{2/3} 
(1+\epsilon_1\phi^r),
\qquad
v\sim \frac{2a(\phi_x)^2}{9\phi^{4/3}} 
(1+\epsilon_2\phi^r),
$$ 
into (\ref{sys}) 
and consider the leading order linear terms in $\epsilon_1$ and 
$\epsilon_2$. This gives a $2\times 2$ homogeneous linear 
system for the $\epsilon_j$, and the vanishing of the 
determinant of the associated 
matrix yields the polynomial equation  
$$ 
9r^4-9r^3-10r^2+8r=0, 
$$ 
whose roots are the resonance values  
\beq \label{reso} 
r=-1,\, 0,\, 2/3,\, 4/3. 
\eeq 
The standard resonance $r=-1$ corresponds to the arbitrary choice of 
$\phi$, while $r=0$ corresponds to $a$ as in   
(\ref{prinb}). From the fact that the other two resonances are also 
non-negative, 
it follows that this expansion 
corresponds to a principal balance, and the fact that 
these other resonances are non-integer rational numbers 
means that the algebraic branching cannot 
be removed simply by changes of the dependent variables (e.g. choosing 
$v^3$ as a new variable does not remove the cube root 
branching from the solution). 

The resonance conditions at the values in (\ref{reso}) 
are all satisfied, which 
means that the leading order terms 
(\ref{prinb}) can be consistently extended to yield the Puiseux series 
\beq 
\label{puiseux} 
u= \phi_t/(2\phi_x)+ \sum_{j=0}^\infty u_j \, \phi^{2/3+j},
\qquad
v= \sum_{j=0}^\infty v_j \,\phi^{-4/3+j}, 
\eeq 
where the singular manifold $\phi$ and  the 
$u_0=9v_0/2=a$, $u_2$, $u_4$ are arbitrarily chosen functions of  
$x$ and $t$, and all other coefficients $u_j$ and  
$v_j$ are determined uniquely 
in terms of these four functions and their derivatives.     

The second type of expansion of the solutions $u,v$ of 
(\ref{sys}) around movable singularities on $\phi (x,t)=0$ 
has the leading order 
behaviour 
\beq
\label{prinb2}
u\sim -\frac{2\phi_x^2}{\phi^{2}},
\qquad
v\sim b\phi^{4},
\eeq
where $b=b(x,t)$ is arbitrary. In the degenerate case $b=0$ (corresponding 
to $v=0$) this just reduces to the well known expansion 
for the KdV equation that was first obtained by the WTC method \cite{wtc}. 
This balance has resonance values   
\beq 
\label{reso2} 
r=-1,0,4,6,  
\eeq 
where $r=-1$ is standard, $r=4$ and $r=6$ correspond to arbitrary functions 
that can be introduced as the coefficients of $\phi^2$ and 
$\phi^4$ in the expansion of $u$ (just as in the case of  KdV).  The 
extra resonance 
value $r=0$ corresponds to the arbitrary function $b$ appearing 
at leading order in the expansion of $v$, as in (\ref{prinb2}), and thus 
we see that this is a second principal balance. Also, the resonance 
conditions for this expansion are trivially satisfied, and it leads 
to a standard series of WTC type, namely  
\beq 
\label{wtc}
u= 2 (\log \phi )_{xx}+ \sum_{j=0}^\infty u_j \,\phi^{j},
\qquad
v= \sum_{j=0}^\infty v_j \,\phi^{4+j},
\eeq
where $\phi$ and the coefficients $v_0=b$, $u_2$, $u_4$ may be 
chosen arbitrarily, with all other coefficients being 
uniquely determined by them. 

It is known that for some integrable systems, 
there are transformations of hodograph (or reciprocal) type which can 
be used to transform a PDE with algebraic branching 
into one with the strong Painlev\'e property \cite{hodo, dhh}, but  
it is by no means clear that this can always be done. In fact, 
in the paper \cite{honeed}, one of us made the  
assertion that  
reciprocal transformations of a certain type 
should succeed in removing all algebraic 
branching from the solutions of certain  
energy-dependent Schr\"odinger 
hierarchies based on Lax operators of the form 
\beq
\label{eschro2}
\lop =\partial_x^2 +\sum_{j=0}^N \lambda^j\, u_j. 
\eeq
This assertion was verified for the principal 
balances in such hierarchies, but a more careful 
consideration of non-principal balances suggests 
that the transformations considered in \cite{honeed} 
were actually 
insufficient to remove all branching from the solutions. 
Thus the question as to  whether 
PDEs  such as (\ref{sys}) and the systems 
treated in \cite{honeed} admit reciprocal 
transformations 
which transform them into PDEs with the strong Painlev\'e property 
remains open. In the next section we consider ODE reductions 
of (\ref{sys}) whose general solutions share the same algebraic 
branching as the original PDE system.

\section{Travelling waves and perturbed H\'enon-Heiles}  

\setcounter{equation}{0}

Here we consider the stationary reductions 
of the higher order flows of the 
hierarchy associated with the system of PDEs (\ref{sys}). 
In fact, more generally we will consider travelling wave solutions 
of the higher flows $\partial_{t_{2N+1}}$ 
such that the dependence on $x$ and $t_{2N+1}$ corresponds to waves 
moving with speed $c$, 
so that 
$$ 
u=u(z), \qquad v=v(z), \qquad z=x-ct_{2N+1} 
$$ 
(where the dependence on the other times $t_{2j+1}$ has been 
suppressed). 
In that case, the travelling wave reduction of the $t_{2N+1}$ flow 
takes the form 
\beq \label{trav}  
\begin{array}{rcc} 
\Big(\partial_z^3+2(u\,\partial_z+\partial_z \,u)\Big)\, (f_N+c/2) 
+\frac{1}{2}\Big( v\,\partial_z+\partial_z \,v\Big)\, f_{N-1} & = & 0, 
\\ 
2 \Big( v\,\partial_z+\partial_z \,v\Big)\, (f_{N}+c/2) & = & 0.  
\end{array} 
\eeq 
The stationary flow corresponds to setting $c=0$. 
 

\begin{figure}[ht!]
\centerline{
\scalebox{1.0}{\includegraphics{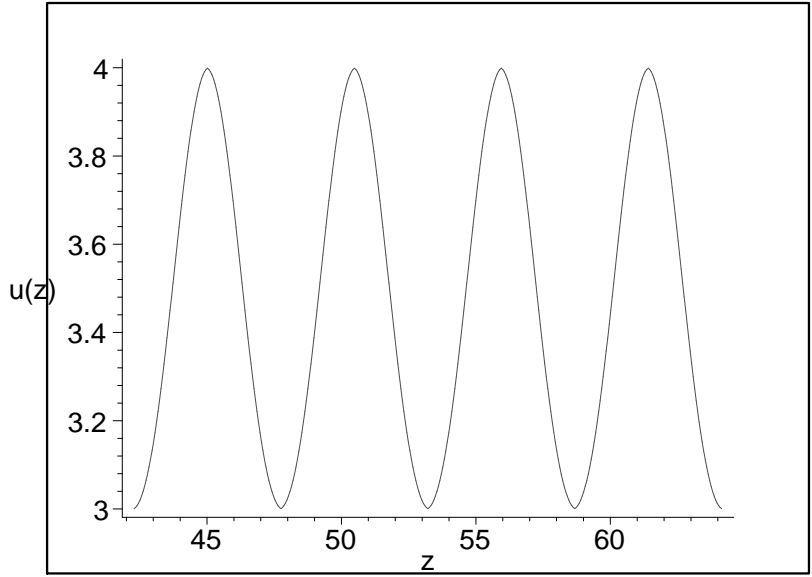}}
}
\caption{ {\it The profile of $u(z)$ for a generic
one-phase solution of the system (\ref{sys}).
} }
\label{onegap}
\end{figure}

Letting $f=f_N+c/2$ and $g=f_{N-1}$, the second 
equation (\ref{trav}) can be integrated 
as 
$$ 
v=\frac{K}{f^2} 
$$ 
for some constant $K$, and then substituting for $v$ in the 
first of these equations and integrating again 
yields the differential equation  
\beq 
ff''-\frac{1}{2}(f')^2+2uf^2+\frac{Kg}{f}+L=0, 
\label{trode} 
\eeq 
where $L$ is another constant. 

First we consider the case $N=1$, which corresponds to the travelling wave 
reduction of the original PDE system (\ref{sys}). 
In that case we have $f_0=g=2$, $f_1=u$ so $f=u+c/2$, and the ODE (\ref{trode}) 
becomes 
\beq \label{1phase} 
ff''-\frac{1}{2}(f')^2+2(f-c/2) f^2
+\frac{2K}{f}+L=0, 
\eeq 
which can be further integrated to yield the quadrature 
\beq 
z-z_0=\int^{u+c/2} \frac{\sqrt{-f}\,\dd f}{\sqrt{2 \mathcal{Q}(f)}}, 
\label{trav3} 
\eeq 
where $\mathcal{Q}$ is the quartic polynomial   
\beq 
\mathcal{Q}(f) =f^4-cf^3-Mf^2-Lf -K\equiv 
(f-e_1) 
(f-e_2)(f-e_3)(f-e_4),   
\label{quartic} 
\eeq 
and $M$, $z_0$ are two more integration constants. 
In the affine coordinates $(f,y)$, it can be seen directly 
that the curve 
\beq 
\label{g2curve} 
fy^2 +2 \mathcal{Q}(f)=0 
\eeq 
defines a two-sheeted cover of the Riemann sphere with six branch 
points, and so has genus two. This can also be seen by 
making the change of variables 
$$ X=-1/f, \qquad Y=y/f^2, $$ 
so that the curve  takes the more standard hyperelliptic 
form 
$$ 
Y^2=2X(e_1X+1)(e_2X+1)(e_3X+1)(e_4X+1) , 
$$ 
and the quadrature (\ref{trav3}) becomes the hyperelliptic integral  
\beq \label{inv} 
z-z_0=\int^{-(u+c/2)^{-1}} \frac{\dd X}{Y}. 
\eeq 

Note that, because the inversion problem (\ref{inv}) 
involves a single integral of one holomorphic differential on a
curve of genus two,  the 
simple ODE (\ref{1phase}) (which can be rewritten in 
Hamiltonian form)  
has deficiency one in the terminology of Abenda and Fedorov 
\cite{abenda, abf}. This can be seen as the origin of the weak 
Kowalevski-Painlev\'e property for the solutions. 
Nonetheless, it is possible to obtain real solutions without any 
singularities on the real axis, 
by choosing suitable initial data to ensure that 
the quantity $f$ lies between two real branch points, $e_3\leq f \leq e_2$ 
say.  To produce numerical plots of such solutions, rather than directly 
evaluating the integral (\ref{trav3}) and then inverting, 
we have found it very convenient to do 
a numerical integration of the first order equation 
$$ 
f'=\pm\sqrt{\frac{-2(f-e_1)(f-e_2)(f-e_3)(f-e_4)}{f} } , 
$$ 
taking into account the sign changes that occur when $f$ reaches one of 
the branch points $e_2,e_3$. We have plotted the profile of $u(z)$ 
for the particular choice $e_1=-2$, $e_2=-3$, $e_3=-4$, $e_4=-5$ 
(see figure \ref{onegap}); 
this corresponds to travelling waves moving with speed 
$c=\sum_j e_j =-14$. We refer to the
travelling wave solutions given by (\ref{trav3}) as ``one-phase''
solutions, by analogy with the corresponding
solutions of KdV and other soliton equations.
For the system (\ref{sys}) it is not clear whether some such 
solutions have a clear interpretation in terms of the 
periodic spectral problem for
the associated energy-dependent Lax operator (\ref{lop}), so the common
terminology ``one-gap'' is perhaps not appropriate here.


\begin{figure}[ht!]
\centerline{
\scalebox{1.0}{\includegraphics{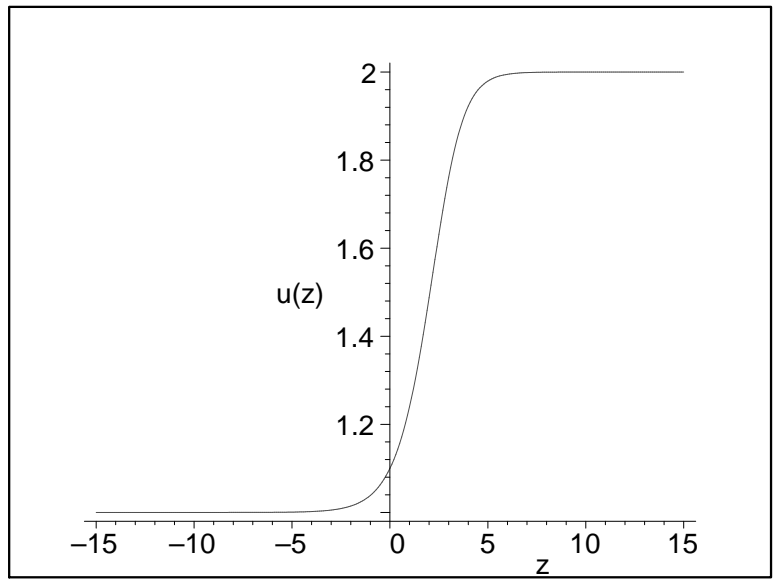}}
}
\caption{ {\it The profile of $u(z)$ for   
a one-soliton solution of the system (\ref{sys}).
} }
\label{onesol}
\end{figure}

In the case when two of the roots of the polynomial $\mathcal{Q}(f)$ 
coincide, the associated hyperelliptic curve 
acquires a singularity and the genus drops 
to one; then the quadrature (\ref{trav3}) can be evaluated 
in terms of elliptic integrals. 
Furthermore, if two pairs of roots coincide (say $e_3=e_1$, 
$e_4=e_2$) then the genus of the curve 
drops to zero,  and then the quadrature (\ref{trav3}) can be 
written explicitly 
as 
\beq 
\label{logform} 
z-z_0= \frac{\pm 1}{\sqrt{2}(k_2^2-k_1^2 )}\left[ 
k_2\, \log \left( 
\frac{k_2-\sqrt{k_1^2+k_2^2-u}}{k_2+\sqrt{k_1^2+k_2^2-u}} \right) 
-k_1 \, \log \left( 
\frac{\sqrt{k_1^2+k_2^2-u}-k_1}{\sqrt{k_1^2+k_2^2-u}+k_1} \right)  
\right], 
\eeq 
where we set  
$$ 
-k_2^2=e_2<e_1=-k_1^2<0. 
$$ 
The above choice of parameters and branches for the logarithms ensures 
that there is a real solution $u=f-c/2$ 
with $e_2<f<e_1<0$,  so that singularities of the type 
$f\sim C\, (z-z_0)^{2/3}$ or $f\sim \tilde{C} \, (z-z_0)^{-2}$ 
are excluded from the real $z$ axis.  
This gives a one-soliton solution for the system (\ref{sys}), which 
has a kink shape (like solitons in the sine-Gordon equation). 
We have plotted an example of such a solution for the choice 
$e_1=-1$, $e_2=-2$, with speed $c=-6$  
(figure \ref{onesol}). 
(Here the term ``one-soliton'' is used in 
a general sense, because as far as we are aware 
the solution does not have a direct interpretation 
in terms of an 
inverse scattering transform.) 

In the case $N=2$ (the $t_5$ flow of the PDE hierarchy) 
we find that the travelling waves 
are equivalent to the Hamiltonian system  with  
two degrees
of freedom defined by the natural Hamiltonian
\beq
h_1=\frac{1}{2}(p_1^2+p_2^2)+q_1^3+\frac{1}{2}q_1q_2^2+2cq_1
-\frac{\ell^2}{2q_2^2}+ \frac{256eq_1}{q_2^4}.
\label{hham}
\eeq
The coordinates $q_j$ and momenta $p_j$ are canonically conjugate,
so $\{ p_j, q_k \}=\delta_{jk}$, 
and Hamilton's equations
\beq
\frac{\dd q_j}{\dd z}=\{ h_1,q_j\},
\qquad
\frac{\dd p_j}{\dd z}=\{ h_1,p_j\},
\qquad
j=1,2
 \label{hameq}
\eeq
are equivalent to the ordinary differential equation 
(\ref{trode}) for
travelling wave solutions of the  
$t_5$ flow upon setting 
\beq \label{link} 
u=g=q_1, \quad v=\frac{4e}{f^2}=\frac{256e}{q_2^4}, 
\quad f=-q_2^2/8=\frac{1}{4}\Big(u''+3u^2+v+2c\Big). 
\eeq 
The derivation of the equations for $q_j$, $p_j$ in Hamiltonian 
form is omitted, since it is essentially the same as the corresponding 
calculation for the stationary flow of the fifth order 
KdV equation carried out by Fordy \cite{fordyhh}; Fordy's results 
are reproduced by setting $e=0$, $c=0$. 
   
The equations of motion (\ref{hameq}) can be written in the
form of a Lax equation
$$
\frac{\dd \LM}{\dd z}=[\M , \LM ],
$$
where the Lax matrix $\LM$ is
\beq
\label{lax}
\LM (\la ; q_j,p_j )=\left(\begin{array}{cc}
-p_1 \la +\frac{1}{4} p_2q_2
&  4\la^2+2q_1\la -\frac{1}{4}q_2^2 \\  
\mathcal{B}(\la  ;q_j,p_j)
 & p_1 \la -\frac{1}{4} p_2q_2
\end{array}\right),
 \eeq
with
$$\mathcal{B}( \la  ;q_j,p_j) = 4\la^3  -2q_1\la^2 +
\left(\frac{1}{4}q_2^2+q_1^2+ 2c\right) \la  
+\frac{1}{4}p_2^2-\frac{\ell^2}{4q_2^2} + \frac{128eq_1}{q_2^4} 
+\frac{16e}{q_2^2\la}.$$
The Lax equation  is the compatibility condition for the
linear system
\beq \label{linear}
\LM \Phi = 2\mu\Phi, \qquad \frac{\dd \Phi }{\dd z}=\M \Phi,
\qquad \M =\left(\begin{array}{cc} 0 & 1 \\
\la - q_1 - \frac{64e}{q_2^4\la}  & 0 \end{array}\right). \eeq
The 
spectral curve given by   
$
\mathrm{det} \, (\LM -2\mu   {\bf 1} )= 0$ 
has the explicit form 
\beq 
\label{specc} 
\mu^2-
4\la^5-2c\la^3-\frac{1}{2}h_1\la^2-
\frac{1}{2}h_2\la -\frac{\ell^2}{64}+ \frac{e}{\la}=0,
\eeq 
with
$$
h_2=\frac{1}{4}q_1p_2^2-\frac{1}{4}q_2p_2p_1-\frac{1}{32}q_2^4
- \frac{1}{8}\left(q_1^2 +2c \right)q_2^2
-\frac{\ell^2q_1}{4q_2^2} +\frac{32e}{q_2^2}+\frac{128eq_1}{q_2^4} 
$$
being the second independent integral, in involution with $h_1$
i.e. $\{h_1,h_2\}=0$. The integral $h_2$ generates a second
commuting flow
$$
\frac{\dd q_j}{\dd y}=\{ h_2,q_j\},
\qquad
\frac{\dd p_j}{\dd y}=\{ h_2,p_j\},
\qquad
j=1,2.
$$  
Using the relations (\ref{link}) for $u$ and $v$, the joint 
solutions $q_j(z,y)$, $p_j(z,y)$  of 
the pair of compatible Hamiltonian systems defined by $h_1$ and $h_2$ 
yield solutions of the PDE system (\ref{sys}) upon identifying 
the independent variables $z=x-ct_5$, $y=4t=4t_3$. 
  
At this point  we should comment on the integrable Hamiltonian system  
defined by  (\ref{hham}). Homogeneous Hamiltonians 
of H\'{e}non-Heiles type were
previously classified by using the usual (strong) Painlev\'e  
test, requiring meromorphic solutions \cite{ctw},  
so it is not surprising that  
extensions of such potentials with the weak Painlev\'e property  
do not appear to have been described in detail before.   
However, for systems with two degrees of freedom,  
an extensive search of natural Hamiltonian systems with a second  
invariant has been performed by  
Hietarinta  \cite{hiet}, and he considered integrable  
cases of such systems with homogeneous cubic potentials, as well as  
various perturbations of such systems.   
The potential (\ref{hham}) with a perturbation of the form  
$q_1/q_2^4$ is a special case of equation (3.2.15) in \cite{hiet},   
and can be obtained by making   
a particular choice of the functions $f$ and $g$ there.

In order to reduce the perturbed H\'enon-Heiles 
Hamiltonian system 
to quadratures, we perform separation of variables, which we 
now describe. The separation of variables for the original three 
integrable cases of the H\'enon-Heiles system (with only cubic and 
harmonic terms in the potential) was obtained in \cite{cgravoson} 
(see also \cite{blaszak}). The  
generalized integrable cases (i) and (iii), which Fordy had 
shown to be reductions of the Sawada-Kotera and Kaup-Kupershmidt 
equations respectively \cite{fordy}, 
were only separated quite recently 
\cite{cmv}. As already mentioned, 
the Hamiltonian (\ref{hham}) that we consider 
here is a perturbation of the integrable case (ii) of the 
H\'enon-Heiles system, and so the separation of variables is 
straightforward.    
A new set of canonically conjugate  coordinates and momenta  
$\la_j$, $\pi_j$ can be obtained according to a well known ansatz (see 
\cite{sklyanin}) whereby the coordinates $\la_1,\la_2$ are given by the 
zeros of the upper right hand entry of the Lax matrix   
(\ref{lax}), so we set 
$$ 
4\la^2 + 2q_1\la - \frac{1}{4}q_2^2=4(\la -\la_1)(\la - \la_2)  
$$  
to find 
\beq 
\la_1+\la_2=-\frac{q_1}{2}, \qquad 
\la_1\la_2=-\frac{q_2^2}{16}. 
\label{coords} 
\eeq

In order to find the canonically conjugate momenta, we can write down the 
generating function 
$$ 
G=-2 
p_1(\la_1+\la_2) 
+ 
4p_2  
\sqrt{-\la_1 \la_2}  
, \qquad 
q_j=\frac{\partial G}{\partial p_j}, 
\quad 
\pi_j=\frac{\partial G}{\partial \la_j}, 
\quad 
j=1,2, 
$$ 
which means that $p_j$ are expressed in terms of the new coordinates 
and momenta as 
\beq 
\left( \begin{array}{c} p_1 \\ p_2 \end{array} \right) 
=\frac{\sqrt{-\la_1\la_2}}{2(\la_1-\la_2)}\left( 
\begin{array}{cc} -\la_1/\sqrt{-\la_1\la_2} & \la_2/\sqrt{-\la_1\la_2} \\ 
1 & -1 \end{array} \right) \left( 
\begin{array}{c} \pi_1 \\ \pi_2 \end{array} \right). 
\label{moms} 
\eeq 
Using the above formulae, we can also rewrite $\pi_1$ and $\pi_2$ 
in terms of the original variables $q_j$, $p_j$, but this 
will not be necessary for what follows.

Once we are equipped with the expressions (\ref{coords}) and (\ref{moms}), 
we can rewrite the two Hamiltonian 
functions $h_1$ and $h_2$ in terms of the new variables 
$\la_j$, $\pi_j$. 
Then one can eliminate between 
these two equations to obtain two separated equations 
 \beq 
\frac{\la_j^2\pi_j^2}{16}-
4\la_j^5-2c\la_j^3-\frac{1}{2}h_1\la_j^2-
\frac{1}{2}h_2\la_j -\frac{\ell^2}{64}+ \frac{e}{\la_j}=0, 
\quad j=1,2.  
\label{sep} 
\eeq 
Upon comparing the equation (\ref{specc}) for the 
spectral curve with the two separated equations 
(\ref{sep}), then with $\pi_j=4\mu_j/\la_j$  
the pair of points $(\la_j , \mu_j)$ for $j=1,2$    
lie on the spectral curve. 
Now we can write down the action $S(h_1,h_2;\la_1,\la_2)$ such 
that 
$$ 
\pi_j =   \frac{\partial S}{\partial \la _j},
\qquad
z_j=\frac{\partial S}{\partial h_j},
\qquad
j=1,2,
$$
by setting 
\beq 
S=\int^{\la_1}\frac{4\mu d\la }{\la} 
+ \int^{\la_2}\frac{4\mu d\la }{\la} 
. 
\label{action} 
\eeq 
This yields the quadratures 
\beq \label{z12} 
z_1=\int^{\la_1}\frac{\la \, d\la }{\mu }
+ \int^{\la_2}\frac{\la \,  d\la }{\mu},  
\qquad z_2=\int^{\la_1}\frac{ d\la }{\mu }
+ \int^{\la_2}\frac{  d\la }{\mu},  
\eeq  
with 
$$ 
z_1=z + \,\mathrm{constant}, \qquad 
z_2=y + \,\mathrm{constant}. 
$$ 


\begin{figure}[ht!]
\centerline{
\scalebox{1.0}{\includegraphics{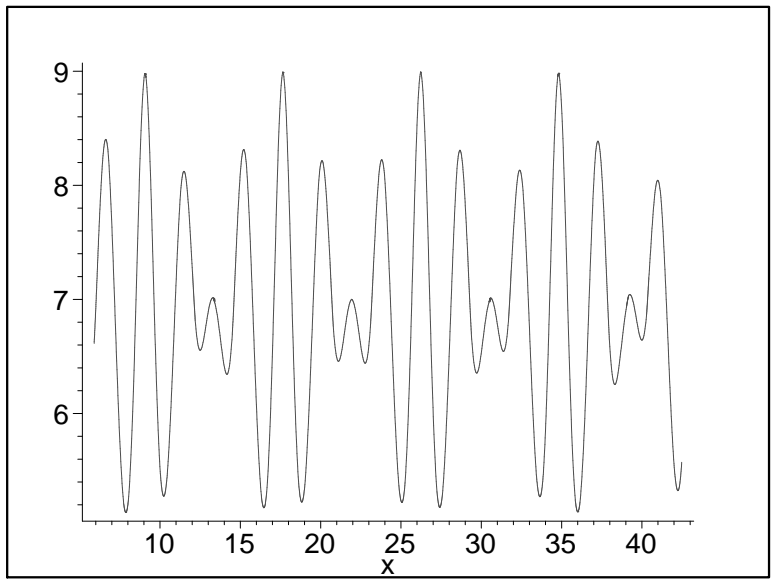}}
}
\caption{ {\it The generic shape of the profile $u(x)=u(x,T)$ for a 
two-phase solution of the system (\ref{sys}), at fixed time $T$.
} }
\label{utwogap}
\end{figure}

The inversion problem (\ref{z12}) associated with 
the hyperelliptic 
curve (\ref{specc}) of genus three involves only two out of 
three independent holomorphic 
differentials. This can be augmented with the corresponding 
relation in terms of a third holomorphic differential, 
$$ 
z_3 = \int^{\la_1}\frac{ d\la }{\la\mu }
+ \int^{\la_2}\frac{  d\la }{\la\mu},
$$ 
and then the extra variable is a function of the other two, 
i.e. $z_3=z_3(z_1,z_2)$. So this perturbed H\'enon-Heiles system 
is another example of an integrable system with deficiency 
one, and instead of  
the standard Jacobi inversion 
involving the threefold symmetric product 
of the curve, one has the twofold product corresponding to  
pairs of points $(\la_1,\mu_1)$, $(\la_2,\mu_2)$. The image 
$(z_1,z_2,z_3)$ of a pair of points under the Abel map lies  
on a stratum in the Jacobian of the curve, and this 
leads to an explanation for the algebraic branching in 
the solutions \cite{abenda}.

The solutions of the perturbed H\'enon-Heiles system produce 
two-phase solutions of the PDE system (\ref{sys}), which are 
the analogues of the two-gap solutions for KdV. The solutions 
corresponding to higher stationary 
flows (and travelling waves) of the PDE hierarchy 
are included in a general ansatz for $N$-phase 
solutions (see \cite{shabat}, for instance). One can set 
$$
\hat{\PS} =\psi \psi^\dagger = \prod_{j=1}^N (\la - \la_j),  
$$
where $\psi$, $\psi^\dagger$ are two independent solutions 
of the Schr\"odinger equation
(\ref{schrod}) whose Wronskian is 
$$\psi_x \psi^\dagger -\psi \psi^\dagger_x=\mu,$$ 
with $\mu = \mu(\la )$ being given by the equation for the 
spectral curve, 
\beq 
\label{Ncurv} 
 \mu^2=4\la^{2N+1} + d\la^{2N}+\ldots -e/\la , \eeq  
which is of genus $N+1$. 
Then $\hat{\PS}$ satisfies the Ermakov-Pinney equation 
\beq
\label{ephat}
\hat{\PS}\hat{\PS}_{xx}-\frac{1}{2}\hat{\PS}_x^2+2W\hat{\PS}^2 + 
\frac{\mu(\la )^2}{2} =0.
\eeq
The expressions for $u$ and $v$ in terms of $\la_j$ (trace formulae) are 
\beq \label{tracef} 
u=-d/4-2\sum_{j=1}^N \la_j, \qquad v=\frac{e}{\prod_{j=1}^N\la_j^2}; 
\eeq  
these follow immediately from expanding (\ref{ephat}) around $\la =\infty$ 
and $\la = 0$. Similarly, expanding around $\la = \la_j$ yields 
the Dubrovin equations 
\beq \label{dubrov}  
\la_j'=\frac{\mu (\la_j)}{\prod_{k\neq j} (\la_j-\la_k)}, 
\qquad j=1,\ldots,N, 
\eeq 
with 
prime denoting derivative with respect to $x$ and 
$\mu$ being a square root of the right hand side of (\ref{Ncurv}).  


\begin{figure}[ht!]
\centerline{
\scalebox{1.0}{\includegraphics{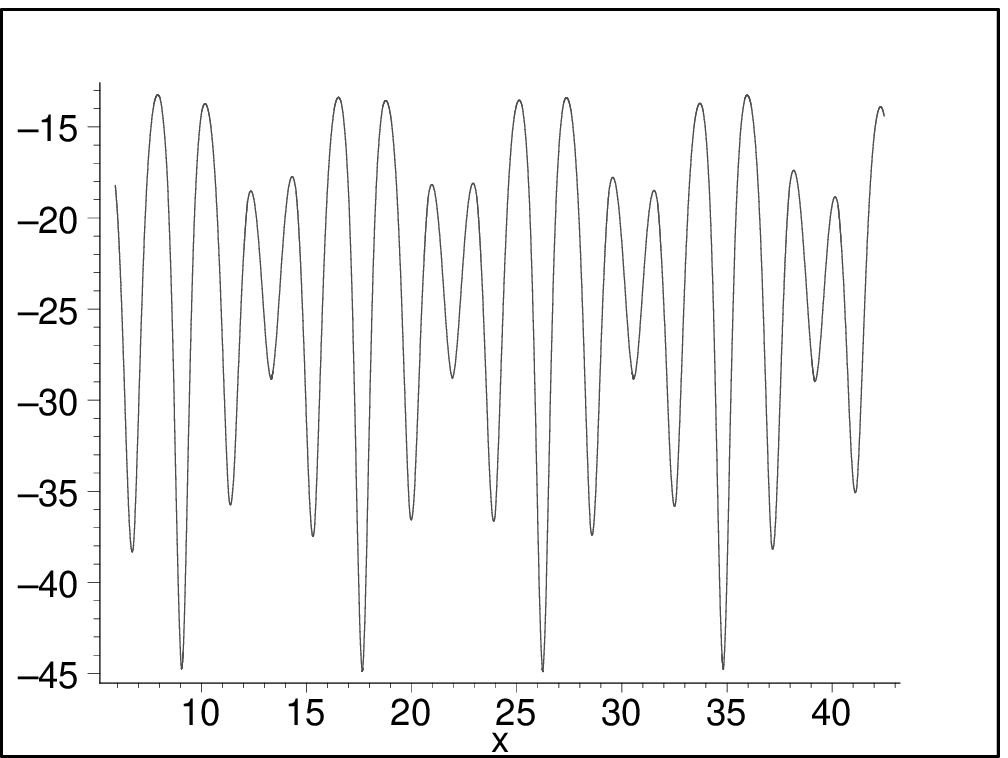}}
}
\caption{ {\it The generic shape of the profile $v(x)=v(x,T)$ for a 
two-phase solution of the system (\ref{sys}), at fixed time $T$.
} }
\label{vtwogap}
\end{figure}

Imposing the time dependence (\ref{psit}) upon $\psi$ and $\psi^\dagger$ 
means that $\hat{\PS}$ evolves with $t$ according to 
$$ 
\hat{\PS}_t=(2u+4\la )\hat{\PS}_x- 2u_x\hat{\PS}, 
$$ 
and this implies that the time evolution of the roots of the 
polynomial $\hat{\PS}$ is given by 
\beq 
\label{lat} 
\la_{j,t}=-\frac{(d/2 + 4 \sum_{k\neq j}\la_k )\, \mu(\la_j)} 
{\prod_{k\neq j} (\la_j-\la_k)}, 
\qquad j=1,\ldots,N  
. 
\eeq 
Just as in the one-phase case, 
the Dubrovin equations are extremely convenient for 
obtaining the solutions numerically. 
We plotted the profile $u(x,t)$ 
of a two-phase solution at a fixed time $t=T$ by 
doing a numerical integration of the equations (\ref{dubrov}) 
for $N=2$ (see figure \ref{utwogap}); for this example 
we placed the branch points of the associated 
genus three curve at $\la =1,2,3,4,5,6$ (as well as 
the fixed branch points at $\la =0,\infty$). The corresponding 
two-phase profile of $v(x,T)$ appears in figure \ref{vtwogap}. 
 

\begin{figure}[ht!]
\centerline{
\scalebox{0.8}{\includegraphics{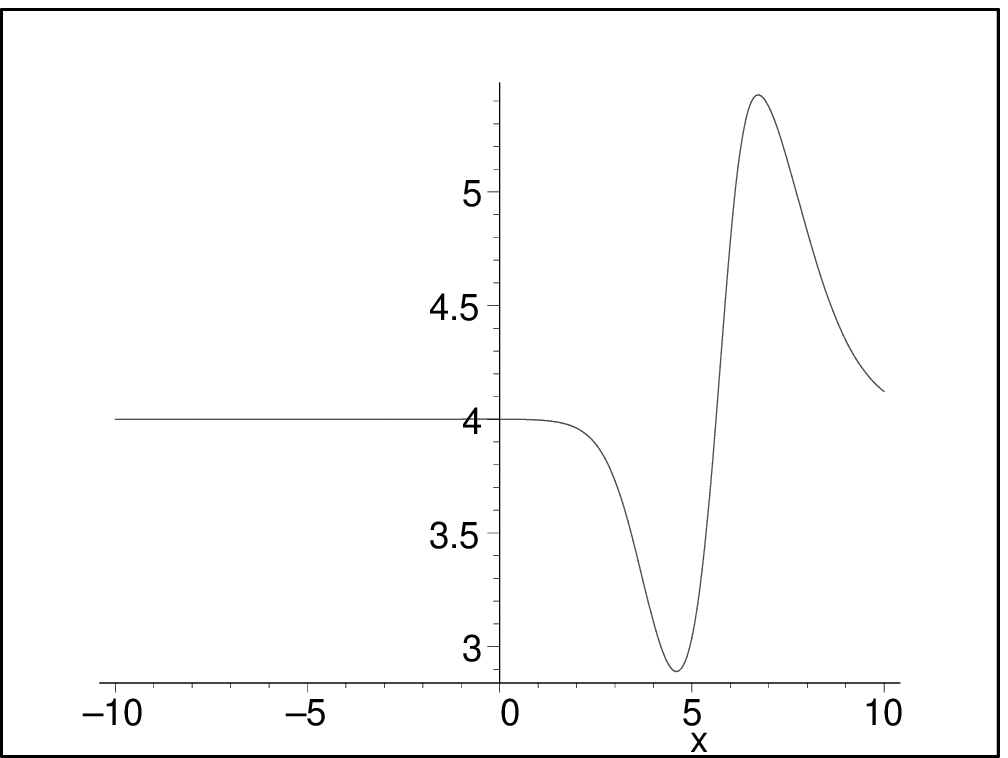}}
}
\caption{ {\it
The profile of $u(x)=u(x,T_0)$ for a
two-soliton solution of the system (\ref{sys}) at
time $T_0$ (before collision).
} }
\label{before}
\end{figure}


\begin{figure}[ht!]
\centerline{
\scalebox{0.8}{\includegraphics{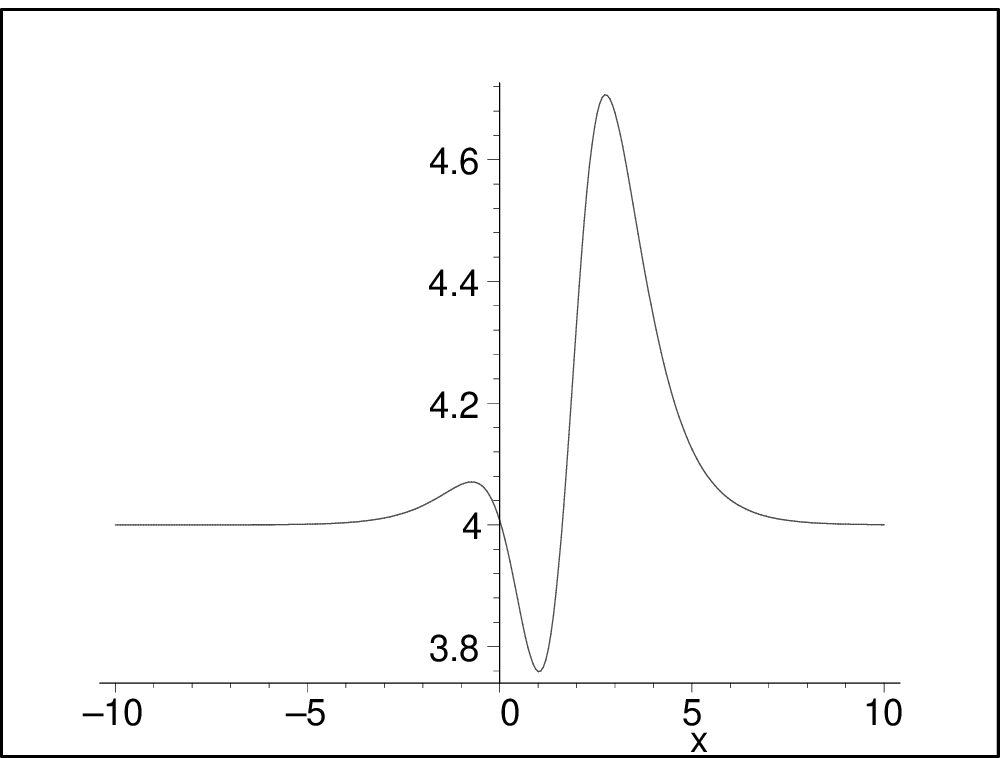}}
}
\caption{ {\it
The profile of $u(x)=u(x,T_1)$ for a
two-soliton solution of the system (\ref{sys}) at
time $T_1$ (just before collision).
} }
\label{justbefore}  
\end{figure}


\begin{figure}[ht!]
\centerline{
\scalebox{0.8}{\includegraphics{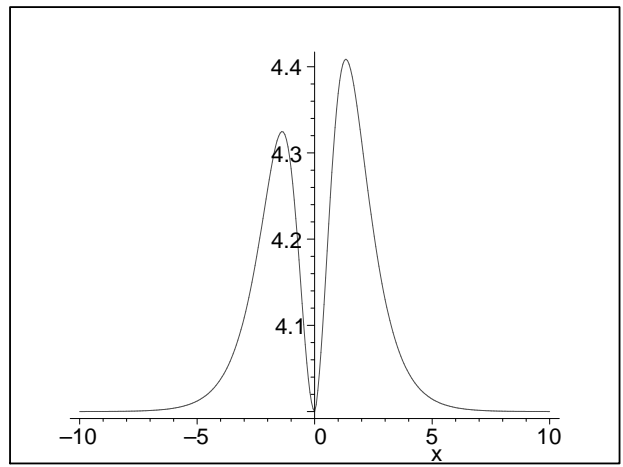}}  
}
\caption{ {\it 
The profile of $u(x)=u(x,T_2)$ for a 
two-soliton solution of the system (\ref{sys}) at 
time $T_2$ (during interaction).
} }
\label{during}
\end{figure}

\begin{figure}[ht!]
\centerline{
\scalebox{0.8}{\includegraphics{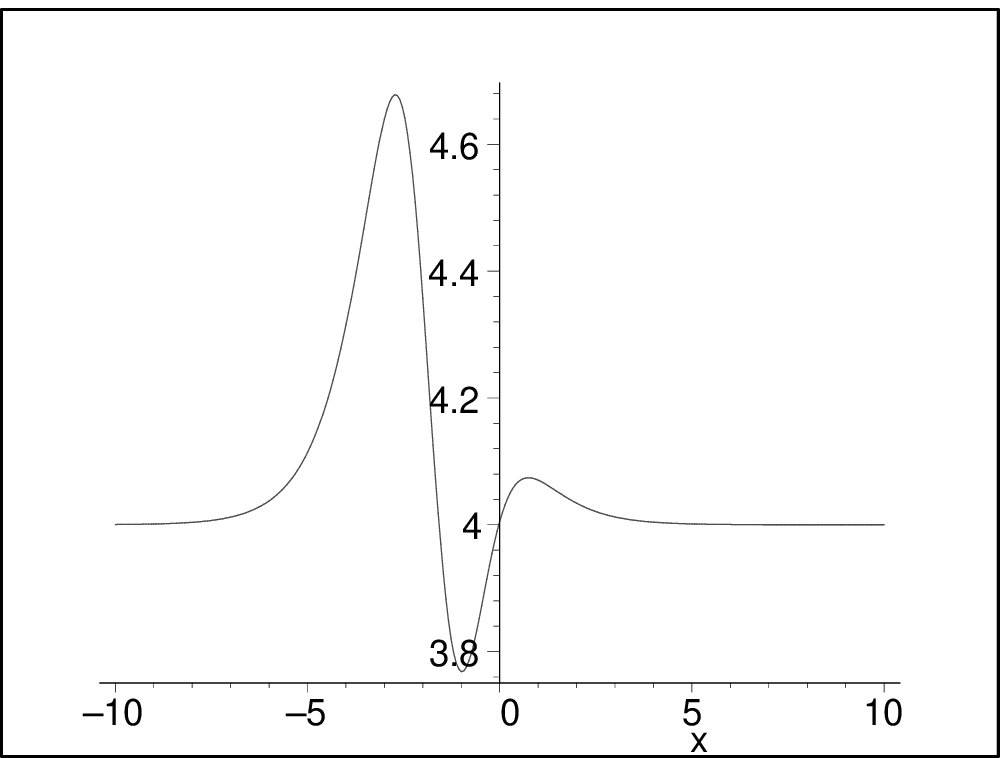}} 
}
\caption{ {\it 
The profile of $u(x)=u(x,T_3)$ for a 
two-soliton solution of the system (\ref{sys}) at 
time $T_3$ (just after collision).
} }
\label{justafter}
\end{figure}

\begin{figure}[ht!]
\centerline{
\scalebox{0.8}{\includegraphics{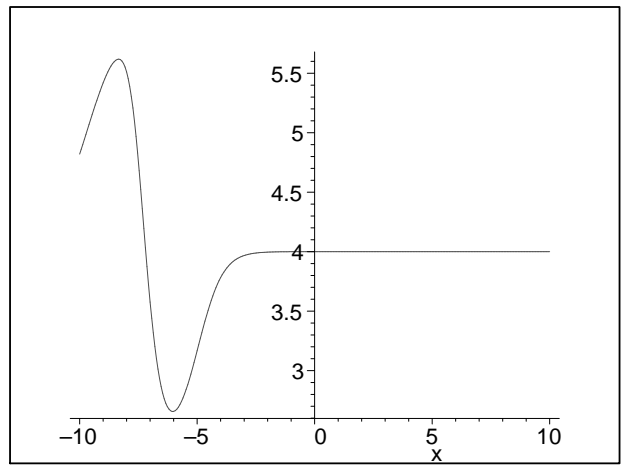}} 
}
\caption{ {\it The profile of $u(x)=u(x,T_4)$ for a
two-soliton solution of the system (\ref{sys}) at
a later time $T_4$ (after interaction).
} }
\label{after}
\end{figure}

\begin{figure}[ht!]
\centerline{
\scalebox{0.8}{\includegraphics{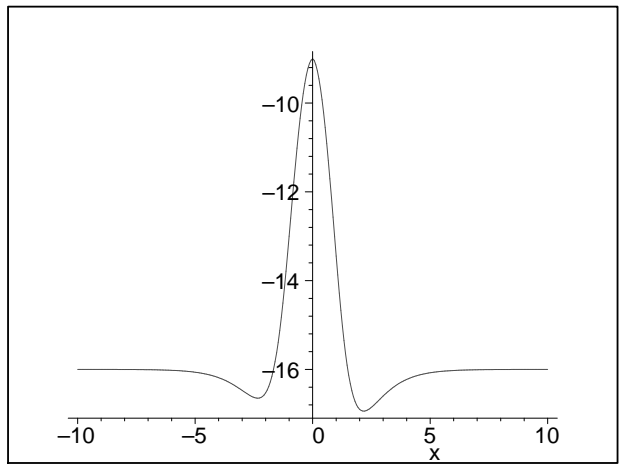}}
}
\caption{ {\it The profile of $v(x)=v(x,T_2)$ for a
two-soliton solution of the system (\ref{sys}) at
time $T_2$ (during interaction).
} }
\label{vduring}
\end{figure}

It is also interesting to consider the two-soliton solutions of (\ref{sys})  
which arise from the coalescence of three pairs of branch points on 
the genus three curve. In order to produce plots of such a solution, 
we performed numerical  integration of the Dubrovin equations 
(\ref{dubrov}) for $N=2$, in the particular case when the right hand side 
of (\ref{Ncurv}) has double roots at $\la = 1,2,3$, using 
particular initial conditions $\la_1|_{x=0}=\la_1 (0,T_0)$, 
$\la_2|_{x=0}=\la_2 (0,T_0)$, with 
$$ 
1<\la_1 (0,T_0)<2, \qquad 
2<\la_2 (0,T_0)<3.  
$$   
This produced the two-soliton profile of 
$u(x,T_0)=-2\la_1(x,T_0)-2\la_2(x,T_0)-d/4$ with $d=-48$, corresponding 
to the configuration of two solitons at some fixed time $t=T_0$ (plotted 
in figure \ref{before}). In order to observe the subsequent 
evolution of the pair of solitons at times $t>T_0$, we 
then integrated the two equations (\ref{lat}) for $j=1,2$, 
starting from the initial 
data $la_1|_{t=T_0}=\la_1 (0,T_0)$,  
$la_2|_{t=0}=\la_1 (0,T_0)$, in order to get values for 
$\la_j (0,t)$ for $t>T_0$. For a fixed sequence of times $t=T_n$, 
the values $\la_j(0,T_n)$, $j=1,2$ were then used as initial 
data for the Dubrovin equations (\ref{dubrov}), in order 
to produce the 
sequence of profiles of $u(x,T_n)$. The interaction between    
the two solitons can be seen in figures \ref{justbefore} - 
\ref{after}. The shape of the field $v(x,t)=e/(\la_1^2\la_2^2)$ 
for the two-soliton solution
(with $e=-144$ in this example)  
is qualitatively similar to that of $u(x,t)$, but for completeness 
we have also plotted $v(x,T_2)$ in figure \ref{vduring}, which is 
at the same stage of the soliton interaction as the 
corresponding profile of $u$ in figure \ref{during}.


\section{Conclusions} 

\setcounter{equation}{0}

The Wahlquist-Estabrook method has proved to be effective for 
finding the Lax pair for the PDE (\ref{tteq}), leading to a 
$2\times 2$ zero curvature representation. 
The latter is equivalent to a scalar Lax pair involving the 
energy-dependent Schr\"odinger equation (\ref{schrod}).   
The solutions of integrable hierarchies associated with energy-dependent 
Schr\"odinger operators have some interesting features compared 
with those of other integrable systems \cite{alber1, alber2}. 
As far as we are aware, the theory of inverse scattering 
has not been developed for such operators 
(apart from the case where $\lop$ is quadratic in $\la$, 
starting with the work \cite{jm}).

The solutions of such hierarchies generally display the weak 
Painlev\'e property \cite{honeed}, and 
in particular the algebro-geometric solutions 
often correspond to finite-dimensional 
integrable Hamiltonian systems with deficiency \cite{abenda}. 
For the $N$-phase solutions of (\ref{sys}) 
described above, the deficiency 
is one: the genus of the curve (\ref{Ncurv}) is one higher than the 
number of degrees of 
freedom. It is interesting to note that, while many such 
Hamiltonian systems were obtained in \cite{abf} by applying 
Dirac reduction to  algebraically completely integrable 
systems that linearize on a $g$-dimensional Jacobian 
associated with a genus $g$ curve (with the 
number of constraints being the deficiency), 
the systems considered here can be viewed 
instead as deformations where the genus increases from 
$g$ to $g+1$:   the parameter $e$ is the deformation 
parameter, and the spectral curves for the KdV hierarchy 
are recovered in the limit $e\to 0$. In particular, the 
perturbed H\'enon-Heiles system with Hamiltonian  
(\ref{hham}) corresponds to a  
deformation from genus two to genus three. One can obtain 
multi-parameter deformations (with correspondingly greater 
deficiency) by considering the Lax operator  
$$ 
\hat{\lop}=\partial_x^2-\lambda+u +\sum_{j=1}^d v_j \, \lambda^{-j}. 
$$

We have plotted some one- and two-phase solutions of the PDE system 
(\ref{sys}), including their solitonic limiting cases. The 
soliton dynamics has some quite  
different features compared with KdV solitons, and is worthy 
of further study. In the absence of an inverse 
scattering transform for the PDE system (\ref{sys}), 
one can wonder to what extent 
the class of soliton-type or $N$-phase solutions represent the 
general solution. For the case of some integrable PDEs, such as 
the KdV and nonlinear Schr\"odinger equations \cite{grin}, 
it is possible to make local approximations to 
arbitrary periodic solutions using finite-gap ones. 

Since the first draft of this paper was written, 
it has been brought to our attention that the Lax pair  
for the system (\ref{sys})   
was written down by Zakharov some time ago 
\cite{zakharov}.  
Ito found the bi-Hamiltonian 
structure and conservation laws for this system 
rewritten in the variables $u$ and  $V=\sqrt{v}$ \cite{ito}.  
In this form, 
Matsuno studied the dispersionless limit and also gave an implicit  
expression for the travelling waves of the system, providing a parametric 
formula for the quadrature (\ref{trav3}), 
or equivalently the hyperelliptic integral (\ref{inv}),
in terms of Jacobi elliptic functions 
(see Appendix B in \cite{matsuno}).  
More recently, the system (\ref{sys}) 
has reappeared in a symmetry classification of  
mixed scalar and   
vector evolution equations 
(see \cite{tsuwolf}, 
section 3.2.3). 

Scaling similarity reductions of integrable PDEs commonly yield 
non-autonomous ODEs of Painlev\'e type, and it has 
been observed that these often inherit the 
same Hamiltonian structures as their autonomous counterparts (see 
e.g. \cite{nahh} and references). However, for  
scaling similarity reductions of the PDE hierarchy 
associated with (\ref{sys}) this appears not to be the case. For example, 
following the method of Clarkson and Kruskal \cite{clark} 
we see that by setting  
$$u(x,t)=\theta^2 \, U(z), \qquad v(x,t)=\theta^4 \, V(z),$$  
with 
$$ 
z=x\theta (t), \qquad \frac{\dd \theta}{\dd t}=\theta^4  
$$ 
(so that $\theta=(-3t+3t_0)^{-1/3}$), 
(\ref{sys}) is reduced to a coupled system of ODEs, namely 
\beq \label{pred} 
\begin{array}{rcl} 
U''' + 6UU' + V'-zU'-2U & = & 0, \\ 
(2U-z)V'+4(U'-1)V & = & 0, 
\end{array} 
\eeq 
where the prime denotes $\dd /\dd z$. 
The integrating factors which allow the ODEs for travelling waves to 
be written as the single second order equation (\ref{1phase}) 
no longer work for this non-autonomous system, except 
in the special case $V\equiv 0$ when (\ref{pred}) 
reduces to the equation PXXXIV (see \cite{ince}, chapter XIV). 
Aside from this special case, the general solution of the fourth 
order system (\ref{pred}) 
will have the weak Painlev\'e property. Nevertheless, 
by applying the technique of Flaschka and Newell 
\cite{fla}, we find that putting  
$$ 
\phi (z;\zeta )=\theta^2 \psi(x,t;\la ), \qquad   
\zeta = \theta^{-2}\la 
$$ 
leads to the linear system 
\beq \label{philin} 
\begin{array}{rcl} 
\phi '' +( U-\zeta +V/(4\zeta) )\, \phi & = & 0, 
\\ 
\zeta \partial_\zeta \phi & = & -(2U-z+4\zeta )\phi' 
+ (U' -2) \phi . 
\end{array} 
\eeq 
The fourth order ODE system arises as the compatibility condition 
for this linear system, so while (\ref{pred}) 
does not have the 
Painlev\'e property it should still 
be possible to interpret it via isomonodromic deformation 
of the second equation (\ref{philin}). 

Another integrable scalar equation of the type 
(\ref{type})  considered in \cite{mnw} 
is equivalent to the system  
\begin{eqnarray}
\label{sys:exkk}
u_t&=&u_{5x}+25u_xu_{xx}+10uu_{3x}+20u^2u_x+v_x,\\
v_t&=&3u_{3x}v+u_{xx}v_x+24uu_xv+4u^2v_x.\nonumber
\end{eqnarray}
The above system is a rewriting of 
equation number (101)
in \cite{mnw},  which is  
obtained by taking $u=w_x$ and eliminating $v$. 
This system reduces to the Kaup-Kupershmidt equation when $v=0$.  
By analogy with our results for the  
system (\ref{tteq}), related to an energy-dependent
operator of second order,  
it is natural to guess that 
the system (\ref{sys:exkk}) should 
possesses a scalar Lax pair with an  
energy-dependent Lax operator of third order,  
which generalizes the Lax pair of the Kaup-Kupershmidt equation. 
Indeed this turns out to be the case, and we obtain the Lax pair  
\begin{eqnarray}
\frac{1}{\lambda}\mathcal{L}\psi &\equiv&\left(\partial_x^3+2u\partial_x+u_x-\frac{v}{9\lambda}-\lambda\right)\psi=0,\\
\partial_t\psi&=& 
-9\lambda\psi_{xx}+(u_{xx}+4u^2)\psi_x-(u_{3x}+8uu_x+12u)\psi  
\end{eqnarray}
for (\ref{sys:exkk}). 
This Lax pair is one member of a family classified by 
Antonowicz, Fordy and Liu \cite{afliu}. 
It is also equivalent to the $3\times 3$ 
zero curvature equation recently given in 
\cite{5lax}. By a slight extension of 
Fordy's results in \cite{fordyhh}, we find that the 
travelling wave reduction of (\ref{sys:exkk}) corresponds to 
another perturbed H\'enon-Heiles system, with 
Hamiltonian 
\beq \label{pertiii} 
h= \frac{1}{2}(p_1^2+p_2^2)+\frac{4}{3}q_1^3+\frac{1}{4}q_1q_2^2+2cq_1
-\frac{\ell^2}{2q_2^2}+ \frac{\tilde{e}}{q_2^6}.
\eeq 
This reduces to the (generalized) integrable case (iii) 
H\'enon-Heiles system when $\tilde{e}=0$, for 
$\tilde{e}\neq 0$ it has the weak Painlev\'e property. 
The possibility of  including the extra term $1/q_2^6$ in the potential, 
while still preserving integrability, was noted some time ago by 
Hietarinta \cite{hiet}. However, at present we do not know 
how to carry out separation of variables for this 
perturbed Hamiltonian system; the results of \cite{cmv} 
cannot be directly extended to this case.

\vspace{.1in} 
\small 
\noindent {\bf Acknowledgements.} Andrew Hone acknowledges 
the support of the EPSRC, and Vladimir Novikov is  
grateful for support from the GIFT project and from an 
EPSRC Postdoctoral Fellowship. Caroline 
Verhoeven is grateful for FWO support in the form of a 
Postdoctoral Research 
Fellowship. Andrew Hone is grateful to Yuri Fedorov 
and Carles Sim\'{o} for useful remarks during his 
visit to Barcelona.    
The authors would also like to thank Jarmo Hietarinta, 
Takayuki Tsuchida and 
Sergei Sakovich for 
valuable comments on the original version of this paper.


\begin{thebibliography}{99}

\bibitem{abenda}Abenda S and Fedorov Y 2000 
{\it  Acta
Appl.
Math.} {\bf 60} 137. 

\bibitem{abf}Abenda S  and Fedorov Y 2001  
{\it J. Nonlin. Math. Phys.} {\bf  8} Supplement 1. 

\bibitem{ars} Ablowitz M J, Ramani A and Segur H 1978  
{\it Lett. Nuovo Cim.} {\bf 23} 333. 

\bibitem{alber1}Alber M S, Luther G G and Marsden J E 1997   
{\it Nonlinearity} {\bf 10} 223.

\bibitem{alber2}Alber M S, Camassa R, Fedorov Yu, Holm DD 
and Marsden J E 1999   
{\it Phys. Lett. A} {\bf 264} 171. 
 
\bibitem{af0} Antonowicz A and Fordy A P 1987
{\it Phys. Lett. A} {\bf 122} 95. 

\bibitem{af1} Antonowicz A and Fordy A P 1987
{\it Physica D} {\bf  28} 345. 


\bibitem{af} Antonowicz A and Fordy A P 1989 
{\it Commun. Math. Phys.} {\bf 124} 465. 

\bibitem{afliu} Antonowicz A, Fordy A P and Liu Q P 1991 
{\it Nonlinearity} {\bf 4} 669. 

\bibitem{blaszak}Blaszak M 1998 
{\it Multi-Hamiltonian theory of dynamical systems}, 
New York: Springer. 

\bibitem{cgravoson}Caboz R, Gavrillov L and Ravoson V 1993 
{\it J. Math Phys. } {\bf 34} 2385. 

\bibitem{ctw}Chang Y F, Tabor M and Weiss J 1982 
{\it J. Math. Phys.} {\bf 23} 531. 


\bibitem{hodo}Clarkson P A, Fokas A S and
Ablowitz M J 1989  
{\it SIAM J. Appl. Math.} {\bf 49} 1188. 

\bibitem{clark}Clarkson P A and Kruskal M D 1989 
{\it J. Math. Phys.} {\bf 30} 2201.

\bibitem{dhh}Degasperis A, Holm D D and Hone A N W 2002  
{\it Theoretical and Mathematical Physics} {\bf  133} 1461. 

\bibitem{erm}Ermakov V P 1880 
{\it Univ. Izv. Kiev Series III} {\bf 9} 1. 

\bibitem{fla}Flaschka H and Newell A C 1980  
{\it  Commun. Math. Phys.} {\bf 76} 65.

\bibitem{fordyhh}Fordy A P 1991  
{\it  Physica} {\bf D 52} 204.

\bibitem{fordy}Fordy A P 1990  {\it Prolongation structures of nonlinear
evolution equations}, in {\it Soliton theory: a survey of results} (ed.
A.P.~Fordy), Manchester University Press,  pp. 403--425.

\bibitem{grin}Grinevich P G 2001 
{\it Physica} {\bf D 152-153} 20. 

\bibitem{hiet}Hietarinta J 1987  
{\it Physics Reports} {\bf 147} 87.  
 
\bibitem{nahh}Hone A N W 1998
{\it Physica} {\bf D 118} 1.

\bibitem{honeed}Hone A N W 1998  
{\it Phys. Lett. A} {\bf 249} 46. 

\bibitem{hone}Hone A N W 1999 
{\it Phys. Lett. A} {\bf 263} 347. 

\bibitem{ince}Ince E L 1926 
{\it Ordinary Differential Equations}. Reprint:
Dover Publications, New York (1956).

\bibitem{ito} 
Ito M 1982 Phys. Lett. A {\bf 91} 335. 

\bibitem{jacobi}Jacobi C G 1842  {\it Vorlesungen \"{u}ber Dynamik},
K\"onigsberg University, ed. A. Clebsch, Reimer, Berlin (1884).

\bibitem{jm}Jaulent M and Miodek I 1976 
{\it Lett. Math. Phys.} {\bf 1} 243. 

\bibitem{kow}Kowalevski S 1889  
{\it Acta Math.} {\it 12} 177;  
Kowalevski S 1890  
{\it Acta Math.} {\bf 14}  81. 


\bibitem{mal}Martinez Alonso L 1980 
{\it J. Math. Phys. } {\bf 21} 2342.  
 
\bibitem{malshabat} Martinez Alonso L and Shabat A B 2002 
{\it Phys. Lett. A} {\bf 299} 359. 

\bibitem{matsuno} 
Matsuno Y 2001 J. Math. Phys.  
{\bf 42} 1744. 
 
\bibitem{miknov} Mikhailov A V and Novikov V S 2002
{\it J. Phys. A: Math. Gen.} {\bf 35}  4775. 

\bibitem{mnw} Mikhailov A V,  Novikov V S and Wang J P 2006 
On classification of integrable non-evolutionary equations, 
{\it Studies in Applied Mathematics} to appear;  
{\tt nlin.SI/0601046}.  

\bibitem{pin}Pinney E 1950 
{\it Proc. Am. Math. Soc. } {\bf 1} 681. 

\bibitem{weak}Ramani A, Dorizzi B and
Grammaticos G 1982  
{\it Phys. Rev. Lett.} {\bf 49} 1538. 

\bibitem{5lax}Sergeyev A 2006 
Zero curvature representation for a new fifth-order integrable system 
{\it preprint} {\tt nlin.SI/0604064}.
 
\bibitem{shabat}Shabat A B 2005 
{\it J. Nonlin. Math. Phys.} {\bf 12}, 
Supplement {\bf 1} 614. 

\bibitem{sklyanin}Sklyanin E V 1995 
Prog. Th. Phys. Supplement {\bf 118} 35. 

\bibitem{tsiganov}Tsiganov A V 1999 
{\it J. Phys. A: Math. Gen.} {\bf 32}  7965. 

\bibitem{tsuwolf} 
Tsuchida T and Wolf T 2005 J. Phys. A: Math. Gen. {\bf 38} 7691.  

\bibitem{cmv} Verhoeven C, Musette M and Conte R 2002 
{\it J. Math. Phys.} {\bf 43} 1906.  

\bibitem{we}Wahlquist H D and Estabrook F B 1975  
{\it J. Math. Phys.} {\bf  16} 1; 
Wahlquist H D and Estabrook F B 1976 
{\it J. Math. Phys.} {\bf  17}
1293. 

\bibitem{wtc}Weiss J, Tabor M and Carnevale J 1983 
{\it J. Math. Phys.} {\bf 24} 522. 

\bibitem{zakharov} 
Zakharov V E 1980  
The inverse scattering  
method, in Solitons,  
Bullough R K and Caudrey P J (eds.),  
Topics in Current Physics vol.  
{\bf 17}, pp. 243--285.  

\end{thebibliography}
\end{document}